\documentclass[aps,prx,showpacs,twocolumn,amsmath,amssymb]{revtex4-1}
\usepackage{graphicx}
\usepackage{dcolumn}
\usepackage{bm}
\usepackage{amsmath}
\usepackage{amsfonts}
\usepackage{amssymb}
\usepackage{multirow}

\usepackage{amsthm}
\usepackage{pinlabel}
\usepackage{natbib}

\usepackage{epstopdf}
\usepackage[abs]{overpic}

\usepackage{cancel}

\usepackage{lipsum}  

\usepackage[normalem]{ulem}

\usepackage[dvipsnames]{xcolor}

\usepackage{braket}

\definecolor{specialgray}{HTML}{505050}
\definecolor{col10K}{HTML}{FFA000}
\definecolor{col300K}{HTML}{924FA4}
\definecolor{colMu}{HTML}{5278BD}
\definecolor{colMuI}{HTML}{924FA4}
\definecolor{newred}{HTML}{D53E4F}
\definecolor{newblue}{HTML}{5278BD}
\definecolor{newcyan}{HTML}{4EBCB3}
\definecolor{newgreen}{HTML}{5CB14E}
\definecolor{newpurple}{HTML}{924FA4}
\definecolor{newyellow}{HTML}{D1C72E}
\definecolor{neworange}{HTML}{D6923C}

\usepackage{lineno}
\usepackage{placeins}

\begin{document}
\title{
Selfconsistent investigation of multichannel superconductivity in a cuprate model system
}
\author{Fabian Schrodi}\email{fabian.schrodi@physics.uu.se}
\author{Alex Aperis}\email{alex.aperis@physics.uu.se}
\author{Peter M. Oppeneer}\email{peter.oppeneer@physics.uu.se}
\affiliation{Department of Physics and Astronomy, Uppsala University, P.\ O.\ Box 516, SE-75120 Uppsala, Sweden}
	
\vskip 0.4cm
\date{\today}

\begin{abstract}
	\noindent The origin of high-temperature superconductivity in cuprates is 
	still an unresolved issue. 
	Among the most likely candidates for mediating the Cooper pair condensate 
	are spin fluctuations and the electron-phonon interaction. 
	While the former have long been proposed to be responsible for various observables in the superconducting state, 
	the latter has recently been shown to produce unconventional gap symmetries 
	when vertex corrections are self-consistently taken into account. Here, we develop multichannel Eliashberg theory 
	to {incorporate} both {pairing} mechanisms.
	Solving self-consistently the full-bandwidth, anisotropic Eliashberg equations for a cuprate model system
	we find the characteristic $d_{x^2 -y^2}$ symmetry of the superconducting gap and a reasonable order of magnitude for $T_c$ ($50-120$ K).
	 We further find that both mechanisms support an unconventional $d$-wave symmetry of the order parameter, 
	 yet the electron-phonon interaction is chiefly responsible for the Cooper pairing and high $T_c$, whereas
	the
	 spin fluctuations 
	 {have} a suppressing effect on the gap magnitude and critical temperature. 
\end{abstract}

\maketitle

\section{Introduction}

After the discovery of the first cuprate superconductor more than 30 years ago \cite{Bednorz1986} many more members of this family have since been found, a large percentage of which exhibit high-temperature superconductivity. To this date, the record holder for $T_c$ ($ \gtrsim 130$ K) at ambient pressure is a copper-based compound \cite{Schilling1993}. Despite an astonishing effort in theoretical and experimental research over several decades, the origin of superconductivity in cuprates is still an unresolved question. 

One of the most controversial aspects in cuprates is the Cooper pairing mechanism, with electron-phonon interaction (EPI) and an electronic mechanism, spin fluctuations (SFs), being the most likely candidates to mediate superconductivity. Among the scientific community purely electronic mechanisms are,  since decades, significantly more popular due to various reasons. First, many undoped cuprates show antiferromagnetic order and become superconducting only upon sufficient electron or hole doping \cite{Taillefer2010,Rybicki2016}, which naturally gives rise to an association of Cooper pairing with magnetic interactions. There is also substantial experimental evidence that the proximity to antiferromagnetic order plays a role for superconductivity, because the magnon excitation spectrum in the superconducting state can be interpreted as softened magnons from the undoped parent phase \cite{LeTacon2011,Guarise2014}. 

Second, the observed unconventional $d$-wave symmetry of the superconducting gap \cite{Chen1994,Shen1995,Kirtley1995,Shi2008} cannot be explained by conventional isotropic EPI. In SF theories, on the other hand, such a sign change of the superconductivity order parameter in the Brillouin zone (BZ) is explicable as the interaction follows approximately the Fermi surface (FS) nesting wave vector. This observation is based on theoretical calculations of SFs and charge fluctuations (CFs) within e.g.\ the random phase approximation (RPA) and fluctuation exchange formalism (FLEX), see Refs.\,\cite{Schmalian1996,Moriya2006,Scalapino2012} and references therein.

On the other hand, ARPES measurements in cuprates have shown that phonons heavily influence the electron dynamics, and therefore might play an important role for superconductivity as well \cite{Lanzara2001}. In addition, other relevant signatures of EPI, such as the isotope effect have been found \cite{Iwasawa2008,Verga2003,Berthod2017,Shen2002,Devereaux2004}. However, as mentioned,
one of the main arguments against EPI being relevant as mediator of superconductivity is the $d$-wave gap symmetry. To resolve this,  
it has been argued
that such an order parameter can be produced from very anisotropic, small momentum transfer electron-phonon scattering {\cite{Kulic2006, Varelogiannis1998}}, which might possibly be realized in cuprate systems due to an unusually large dielectric constant {\cite{Abrikosov1994,Weger1996}}.

Even when considering isotropic EPI, a mechanism that cannot produce unconventional order parameters in conventional `standard' theories, vertex corrections to the EPI can lead to partially negative contributions to the renormalized effective interaction \cite{Schrodi2020_2}, thereby enhancing the coupling strength in the $d$-wave channel \cite{Ishihara2004,Ishihara2006}. Indeed, self-consistent calculations for cuprate model systems have shown that nonadiabatic Eliashberg theory beyond Migdal's approximation produces the experimentally observed $d$-wave order parameter \cite{Hague2006,Schrodi2021}.

It is for these reasons that the mechanism of superconductivity in cuprates is quite controversial, and it is not unlikely that both, EPI and SFs, play an important role in mediating the record critical temperatures. With the current communication we want to contribute to the ongoing discussion by investigating a cuprate model system within a multichannel Eliashberg formalism \cite{Eliashberg1960,Schrodi2020_3}, i.e., we consider both  EPI and SFs in our theory. For the electronic part of the interaction we employ a single-orbital FLEX formalism, while the EPI is treated beyond Migdal's approximation \cite{Migdal1958}, taking into account all first and second order electron-phonon scattering processes \cite{Schrodi2020_2}. With the resulting Eliashberg theory  we are capable of studying the cooperative and competitive effects of both mediators of superconductivity. 

Our self-consistent, anisotropic full-bandwith calculations reveal that SFs and CFs alone do not lead to a stable superconducting solution in the temperature range considered. This stems from the small wave-vector contributions to the interaction, which support a different gap symmetry than the dominant coupling around the nesting wave vector, an effect that also has been observed in Fe-based superconductors \cite{Yamase2020,Schrodi2020_3}. In contrast, EPI beyond Migdal's approximation leads to the expected unconventional $d$-wave pairing, in agreement with earlier studies \cite{Hague2006,Schrodi2021}, and the resulting values for $T_c$ vary according to the electron-phonon coupling strength.

When both interactions are used 
 we find similar superconducting solutions, where an increase in the electron-phonon coupling strength increases $T_c$, while stronger influence of the electronic mechanism leads to a reduction of the
critical temperatures. Further, we observe that due to the EPI-mediated superconductivity, the SF interactions at small wave vectors are decreased, thereby reducing the competition for other symmetries than $d$-wave. The inclusion of vertex corrections beyond Migdal's approximation is essential to obtain the here-presented results, because isotropic EPI as used in standard theories cannot produce unconventional gap symmetries. Our work is a major
step towards realistic multichannel superconductivity calculations in high-$T_c$ materials, in which the interplay of EPI with purely electronic mechanisms can be thoroughly analyzed.

\section{Methodology}

The interacting state of the system is characterized by the electronic self-energy $\hat{\Sigma}_k$ and electron Green's function $\hat{G}_k$, where we use the four-momentum notation $k=(\mathbf{k},i\omega_m)$, $q=(\mathbf{q},iq_l)$, with fermion Matsubara frequencies $\omega_m=\pi T(2m+1)$ and boson frequencies $q_l=2\pi Tl$. Our theory is formulated in Nambu space, spanned by Pauli matrices $\hat{\rho}_i$, $i=0,1,2,3$, allowing us to write the electron self-energy as matrix decomposition
\begin{align}
\hat{\Sigma}_k = i\omega_k(1-Z_k)\hat{\rho}_0 + \chi_k\hat{\rho}_3 + \phi_k\hat{\rho}_1 . \label{sigmaDef}
\end{align}
Above, $Z_k$ is the mass renormalization, $\chi_k$ the chemical potential renormalization, and $\phi_k$ the superconductivity order parameter. 

In the non-interacting state the electron Green's function has the well-known form $\hat{G}_k^{(0)}=[i\omega_k\hat{\rho}_0-\xi_k\hat{\rho}_3]^{-1}$, which, together with Eq.\,(\ref{sigmaDef}), determines the functional form of the Dyson equation $\hat{G}_k=\hat{G}^{(0)}_k + \hat{G}^{(0)}_k\hat{\Sigma}_k\hat{G}_k$ for the fully interacting Green's function. Here $\xi_k$ describes the single-band electron energy. By using the definitions $\gamma_k^{(Z)}=\omega_kZ_k/\Theta_k$, $\gamma_k^{(\chi)}=(\xi_k+\chi_k)/\Theta_k$ and $\gamma_k^{(\phi)}=\phi_k/\Theta_k$, with
\begin{align}
\Theta_k = \big(i\omega_k Z_k\big)^2 - \big(\xi_k+\chi_k\big)^2 - \phi_k^2 ,
\end{align}
the above expressions lead to
\begin{align}
\hat{G}_k = i\gamma_k^{(Z)}\hat{\rho}_0 + \gamma_k^{(\chi)}\hat{\rho}_3 + \gamma_k^{(\phi)} \hat{\rho}_1 .
\end{align}

As mentioned in the Introduction, we take into account EPI beyond Migdal's approximation, as well as SFs and CFs. Accordingly, we can split the electronic self-energy into
\begin{align}
\hat{\Sigma}_k = \hat{\Sigma}_k^{(\mathrm{ep})} + \hat{\Sigma}_k^{(\mathrm{s})} + \hat{\Sigma}_k^{(\mathrm{c})} , \label{sigmaFull}
\end{align}
with labels $(\mathrm{ep})$, $(\mathrm{s)}$ and $(\mathrm{c})$ for EPI, SFs, and CFs, respectively. We first discuss the purely electronic mediators of superconductivity in subsection \ref{scf}, and afterwards give details on $\hat{\Sigma}_k^{(\mathrm{ep})}$ in subsection \ref{epi}. The self-consistent Eliashberg equations arising from Eq.\,(\ref{sigmaFull}) are given in subsection \ref{eliash}.

\subsection{Spin and charge fluctuations}\label{scf}

Assuming spin-singlet Cooper pairs, we treat here the purely electronic parts of the electron self-energy similarly to Ref.\,\cite{Lenck1994}, i.e., within a FLEX formalism. The self-energy contributions to SFs and CFs are calculated via
\begin{align}
\hat{\Sigma}_k^{(s)} &= T\sum_{k_1} V^{(s)}_{k-k_1} \hat{\rho}_0 \hat{G}_{k_1} \hat{\rho}_0 ,\\
\hat{\Sigma}_k^{(c)} &= T\sum_{k_1} V^{(c)}_{k-k_1} \hat{\rho}_3 \hat{G}_{k_1} \hat{\rho}_3 ,
\end{align}
where the interactions are defined in terms of the single-orbital, onsite 
$U$ parameter:
\begin{align}
V^{(s)}_{q} = \frac{3}{2} U^2 X^{(s)}_q  , \\
V^{(c)}_{q} = \frac{1}{2} U^2 X^{(c)}_q  .
\end{align}

The electron spin and charge susceptibilities in the interacting state are calculated from the RPA-like expressions
\begin{align}
X^{(s)}_q &= \frac{X^{(s,0)}_q}{1-UX^{(s,0)}_q} , \\
X^{(c)}_q &= \frac{X^{(c,0)}_q}{1+UX^{(c,0)}_q} ,
\end{align}
while the non-renormalized functions are given in terms of the systems Green's function,
\begin{align}
X^{(s,0)}_q &= -\frac{T}{2}\sum_k \mathrm{Tr} \big\{ \hat{G}_k \hat{\rho}_0 \hat{G}_{k+q} \hat{\rho}_0 \big\} , \label{xs0}\\
X^{(c,0)}_q &= -\frac{T}{2}\sum_k \mathrm{Tr} \big\{ \hat{G}_k \hat{\rho}_3 \hat{G}_{k+q} \hat{\rho}_3 \big\} . \label{xc0}
\end{align}
In the above equations we use the fully interacting Green's function, hence the magnon spectrum is iteratively updated making our calculations fully self-consistent.

\subsection{Electron-phonon interaction} \label{epi}

We treat the coupling between electrons and phonons similar to previous works \cite{,Schrodi2020_2,Schrodi2021}, assuming an isotropic coupling constant $g_0$ and an Einstein phonon mode $\Omega$. With $V^{(\mathrm{ep})}_{k-k_1}=2g_0^2\Omega/(\Omega^2+q_{m-m'}^2)$, the electron-self energy takes the form
\begin{align}
\hat{\Sigma}_k^{(\mathrm{ep})} &= T \sum_{k_1} V^{(\mathrm{ep})}_{k-k_1} \hat{\rho}_3 \hat{G}_{k_1} \hat{\rho}_3 \big(1 + g_0^2\hat{\Gamma}_{k,k_1}\big) ,
\end{align}
where all first and second order scattering processes are taken into account. The vertex is given by
\begin{align}
\hat{\Gamma}_{k,k_1} = \frac{T}{g_0^2} \sum_{k_2} V_{k_1-k_2}^{(\mathrm{ep})} \hat{G}_{k_2}\hat{\rho}_3 \hat{G}_{k_2-k_1+k}\hat{\rho}_3 . \label{vertex}
\end{align}

To reduce the numerical workload we introduce here a non-interacting state approximation for the vertex, which has been shown to be accurate for cuprate systems \cite{Schrodi2021}, see also Appendix \ref{Appendix} for further details. Within this approximation we use
\begin{align}
\Gamma_{k,k_1} \simeq \Gamma_{k,k_1}^{(0)} = \frac{T}{g_0^2} \sum_{k_2} V_{k_1-k_2}^{(\mathrm{ep})} \hat{G}_{k_2}^{(0)} \hat{\rho}_3 \hat{G}_{k_2-k_1+k}^{(0)}\hat{\rho}_3 , \label{nonint}
\end{align}
so that with definitions $\gamma_k^{(\xi)}=\xi_k/\theta_k$, $\gamma_k^{(\omega)}=\omega_k/\theta_k$ and $\theta_k = -\omega_k^2 - \xi_k^2$ we get
\begin{align}
\hat{\Gamma}_{k,k_1}^{(0)} =& \frac{T}{g_0^2} \sum_{k_2} V^{(\mathrm{ep})}_{k_1-k_2} \big[ \big(\gamma_{k_2}^{(\xi)}\gamma_{k_2-k_1+k}^{(\xi)} - \gamma_{k_2}^{(\omega)}\gamma_{k_2-k_1+k}^{(\omega)} \big) \hat{\rho}_0  \nonumber \\
&~~~~~~+ \big(\gamma_{k_2}^{(\xi)}\gamma_{k_2-k_1+k}^{(\omega)}  + \gamma_{k_2}^{(\omega)}\gamma_{k_2-k_1+k}^{(\xi)}\big) i\hat{\rho}_3  \big] . \label{gamma0}
\end{align}
Note, that $\hat{\Gamma}_{k,k_1}^{(0)}$ can be pre-computed and hence, this approach offers a significant numerical advantage.

The here-made truncation of the infinite series of Feynman diagrams for the electron-phonon interaction after the second-order scattering terms is valid for weakly non-adiabatic systems. This is when the non-adiabaticity ratio \cite{Migdal1958}, $\alpha = \Omega/\varepsilon_F$, where $\Omega$ is the energy scale of the phonons and $\varepsilon_F$ that of the electrons, is small,
typically of the order $\alpha = 0.1 - 0.2$ \cite{Schrodi2020_2}. 

\begin{figure*}[t!]
	\centering
	\includegraphics[width=1\textwidth]{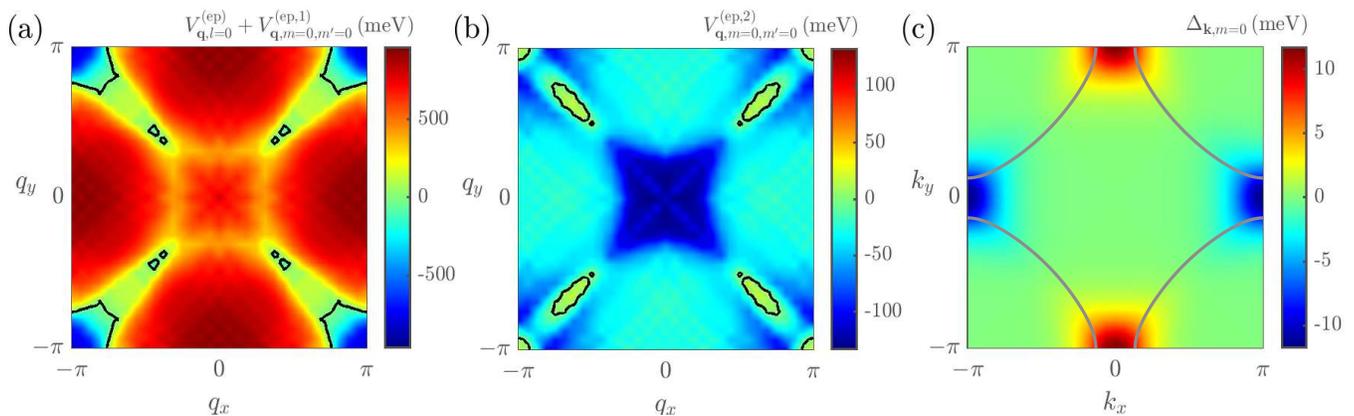}
	\caption{(a) Electron-phonon interaction kernels $V^{(\mathrm{ep})}_{\mathbf{q},m-m'}+V^{(\mathrm{ep},1)}_{\mathbf{q},m,m'}$ at $m=m'=0$, calculated for $g_0=170\,\mathrm{meV}$ and $T=80\,\mathrm{K}$. (b) Same as (a), but for the {second}
	electron-phonon kernel $V^{(\mathrm{ep},2)}_{\mathbf{q},m,m'}$. Solid black lines indicate zero crossings of the kernel functions. (c) Self-consistent zero-frequency component of the superconducting gap, having $d_{x^2-y^2}$-symmetry, obtained for $g_0=170\,\mathrm{meV}$, $T=80\,\mathrm{K}$ and $U=0\,\mathrm{eV}$. The cuprate Fermi surface is shown in gray color.}	\label{gapexample}
\end{figure*}

\subsection{Eliashberg equations}\label{eliash}

By using the standard procedure we derive the Eliashberg equations from the two expressions for the self-energy Eqs.\,(\ref{sigmaDef}) and (\ref{sigmaFull}), and find
\begin{align}
Z_k &= 1 - \frac{T}{\omega_k} \sum_{k_1} \Big[ \big( V^{(+)}_{k-k_1} +  V_{k,k_1}^{(\mathrm{ep},1)} \big) \gamma_{k_1}^{(Z)} +   V_{k,k_1}^{(\mathrm{ep},2)}\gamma_{k_1}^{(\chi)} \Big] , \label{z}\\
\chi_k &= T \sum_{k_1} \Big[\big( V^{(+)}_{k-k_1} + V_{k,k_1}^{(\mathrm{ep},1)} \big) \gamma_{k_1}^{(\chi)} - V^{(\mathrm{ep},2)}_{k,k_1}\gamma_{k_1}^{(Z)}  \Big]  , \label{chi}\\
\phi_k &= -T \sum_{k_1} \big(  V^{(-)}_{k-k_1}  + V_{k,k_1}^{(\mathrm{ep},1)} \big) \gamma_{k_1}^{(\phi)}  . \label{phi}
\end{align}
Here the first-order interaction kernels containing EPI, SFs, and CFs are defined as
\begin{align}
V_{k-k_1}^{(\pm)} = V_{k-k_1}^{(\mathrm{ep})} + V_{k-k_1}^{(\mathrm{c})} \pm V_{k-k_1}^{(\mathrm{s})} ,
\end{align}
while second-order electron-phonon scattering within the non-interacting state approximation is given by
\begin{align}
V_{k,k_1}^{(\mathrm{ep},1)} &= T V^{(\mathrm{ep})}_{k-k_1} \sum_{k_2} V^{(\mathrm{ep})}_{k_1-k_2} \nonumber \\
&~~~~ \times \big( \gamma_{k_2}^{(\xi)}\gamma_{k_2-k_1+k}^{(\xi)} -\gamma_{k_2}^{(\omega)}\gamma_{k_2-k_1+k}^{(\omega)} \big) , \label{kern1}\\ 
V_{k,k_1}^{(\mathrm{ep},2)} &= T V^{(\mathrm{ep})}_{k-k_1} \sum_{k_2} V^{(\mathrm{ep})}_{k_1-k_2} \nonumber\\
&~~~~ \times \big( \gamma_{k_2}^{(\xi)}\gamma_{k_2-k_1+k}^{(\omega)}    + \gamma_{k_2}^{(\omega)}\gamma_{k_2-k_1+k}^{(\xi)}  \big) . \label{kern2}
\end{align}
In summary, when solving Eqs.\,(\ref{z})-(\ref{phi}) we include SFs and CFs on the level of the FLEX approach, while EPI is included up to all second-order processes, with a non-interacting state approximation for two of the three electron Green's functions appearing in the vertex correction.

Our theory as presented here is 
a step forward to combining SFs and CFs with phonons in vertex-corrected Eliashberg theory. The Cooper pair-breaking Coulomb interaction is assumed to be already included in 
the electronic energy dispersions {and we omit its effect on the off-diagonal self-energies for simplicity}. It should further be noted that we neglect the phonon self-energy here, which might be important for high-temperature superconductors \cite{Zeyher1990}. {Also,} both mediators of superconductivity are {included but} treated independently of each other, i.e., we do not include a renormalization of the magnon spectrum due to the electron-phonon coupling. These approximations are made for computational accessibility, so as to obtain
quantitative results.

The numerical calculations in this work have been carried out with the Uppsala Superconductivity Code (\textsc{uppsc}) {\cite{UppSC,Aperis2015,Schrodi2019,Schrodi2020_2,Schrodi2021_2}.
}

\section{Results}
\label{Results}

Our model system for electron-doped cuprates is given by the tight-binding description
\begin{align}
\xi_{\mathbf{k}} = t[\cos(k_x)+\cos(k_y)] + t' \cos(k_x)\cos(k_y) - \mu ,
\end{align}
similar to the one used in Ref.\,\cite{Schrodi2021}. The hopping energies and chemical potential are fixed at $t=-0.25\,\mathrm{eV}$, $t'=0.1\,\mathrm{eV}$ and $\mu=-0.07\,\mathrm{eV}$. The resulting Fermi surface is shown in Fig.\,\ref{gapexample}(c) as gray lines. Further, we use an Einstein phonon frequency $\Omega=50\,\mathrm{meV}$ for the EPI, which has been shown to be relevant for cuprate systems \cite{Lanzara2001}. {We have checked that moderate changes in $\Omega$ lead to qualitatively similar results as presented here.} The electron-phonon scattering strength $g_0$ and the onsite 
$U$ are treated as free parameters in the following.

As a first step we focus on the limiting case of $U=0\,\mathrm{eV}$, which serves as a benchmark calculation for our non-interacting state approximation for the vertex function, compare Eq.\,(\ref{gamma0}). For $g_0=170\,\mathrm{meV}$ we show the interaction $V^{(\mathrm{ep})}_{\mathbf{q},l}+V^{(\mathrm{ep,1})}_{\mathbf{q},m,m'}$, which enters each of Eqs.\,(\ref{z})-(\ref{phi}), for $m=m'=0$ and $T=80\,\mathrm{K}$ in Fig.\,\ref{gapexample}(a). In accordance with Ref.\,\cite{Schrodi2021}, this function is repulsive around the nesting wave vector $\mathbf{q}\sim(-\pi,\pi)$, i.e., the M point, and attractive in most parts of the remaining BZ. Such characteristics are favorable for a $d$-wave symmetry of the superconducting gap. In panel (b) of the same figure we show $V^{(\mathrm{ep,2})}_{\mathbf{q},m=0,m'=0}$, which is negative in most parts of the BZ and significantly smaller in magnitude than the interaction in panel (a). It further deserves mentioning that $V^{(\mathrm{ep,2})}_{\mathbf{q},m,m'}$ enters only in Eqs.\,(\ref{z}) and (\ref{chi}), and therefore it should not have a big impact on the superconducting properties. 

When solving the Eliashberg equations self-consistently for this set of parameters we find the experimentally observed $d_{x^2-y^2}$ symmetry of the zero-frequency gap function $\Delta_{\mathbf{k},m=0}=\phi_{\mathbf{k},m=0}/Z_{\mathbf{k},m=0}$, which is shown in Fig.\,\ref{gapexample}(c). This is the only symmetry of the superconducting order parameter found here. Therefore, when discussing superconductivity for this system with any other parameter settings below, everything is to be understood in terms of a $d$-wave symmetric gap function. The chosen value $g_0=170\,\mathrm{meV}$ corresponds to an input electron-phonon coupling strength of $\lambda_0=2$. The effective interaction strength $\lambda_m$ due to EPI can be calculated above $T_c$ \cite{Schrodi2021}, via
\begin{align}
\langle Z_{\mathbf{k},m=0}\rangle_{\mathbf{k_F}}|_{T>T_c} = 1+\lambda_m - \lambda_m^2 \frac{\pi^2\Omega N_0}{4}\frac{3\sinh{\frac{\Omega}{T}}-\frac{\Omega}{T}}{\cosh{\frac{\Omega}{T}-1}} .
\end{align}
At $T=120\,\mathrm{K}>T_c$ we find $\lambda_m=1.5$, which is a significant reduction compared to the input coupling. 
Note, that even more realistic values for the electron-phonon coupling strength \cite{Lanzara2001} are {obtained} when the vertex-corrected Eliashberg equations are solved without any approximation \cite{Schrodi2021}.

The other extreme case would be to only consider spin and charge fluctuations, and no influence of EPI, i.e.\ choosing $g_0=0\,\mathrm{meV}$. For this setting it was not possible to stabilize a self-consistent solution {for the temperatures considered here, $T \ge 30$\,K,} regardless of the choice of $U$ and $T$ (minimum temperature tested was $T=30\,\mathrm{K}$). This result qualitatively agrees with recent studies on Fe-based superconductors \cite{Yamase2020,Schrodi2020_3} and we comment on it further below. 

{To check the consistency of our approach with earlier work \cite{Lenck1994}, we tested smaller temperatures down to $T=6$\,K for the limiting case of vanishing $g_0$. It turns out that a finite superconducting gap with $d_{x^2-y^2}$ symmetry is obtained at temperatures smaller than 15\,K when choosing $U=0.35$\,eV. This choice puts the system in very close vicinity of a magnetic instability, which is reflected in an almost delta-peak-like shape of the SFs kernel. In the limit of $T \rightarrow 0$\,K we {computed} a gap magnitude of 5.4\,meV. These {results}, accompanied by unusually large values of the electron mass renomalization function, are in excellent agreement with previous work \cite{Lenck1994b}.} 

\begin{figure}[tb!]
	\centering
	\includegraphics[width=1\columnwidth]{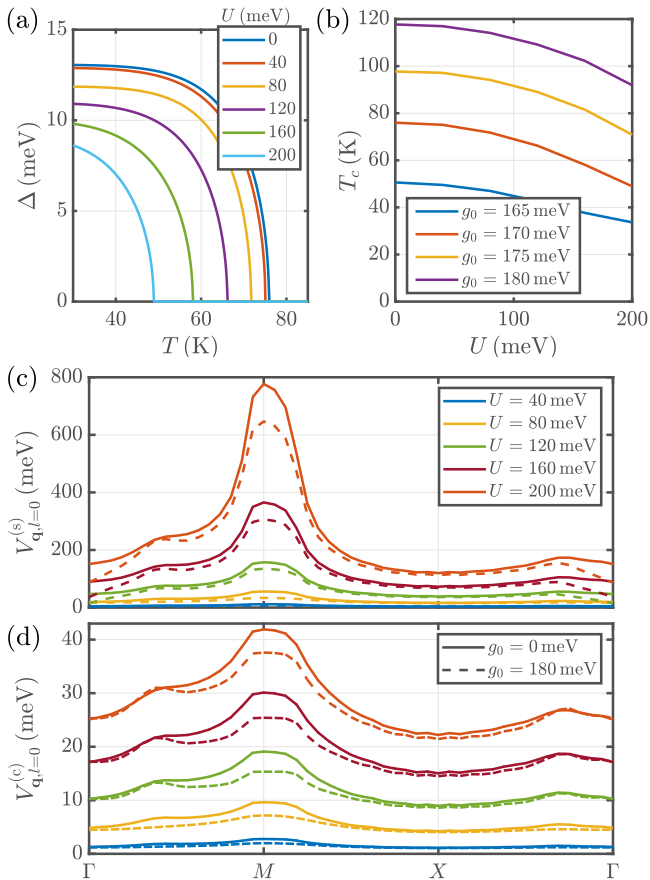}
	\caption{(a) Maximum superconducting gap as function of temperature, calculated for $g_0=170\,\mathrm{meV}$ and different values for $U$ as indicated in the legend. (b) Superconducting critical temperature $T_c$ against $U$ for varying choices of electron-phonon coupling constant $g_0$, see legend. Panels (c) and (d) show the zero-frequency component of the self-consistently obtained spin- and charge-fluctuations kernel, respectively, along high-symmetry lines of the BZ at $T=80\,\mathrm{K}$. Different colors correspond to varying choices of $U$, while solid and dashed lines represent $g_0=0\,\mathrm{meV}$ and $g_0=180\,\mathrm{meV}$, respectively.}
	\label{gaptc}
\end{figure}
To explore the influence of the 
$U$ parameter on the superconducting properties we show in Fig.\,\ref{gaptc}(a) the gap amplitude versus temperature for different choices of $U$ as indicated in the legend, and $g_0=170\,\mathrm{meV}$. As is directly apparent, the gap size and critical temperature decrease upon growing $U$. Following $T_c$ as function of $U$ in panel (b) of the same figure for different values of $g_0$ shows that this behavior is robust against changes in the electron-phonon coupling, and that therefore, 
{for the current level of approximation,} 
spin/charge fluctuations contribute destructively to the Cooper pairing.

To understand these findings we need to examine the self-consistently calculated spin and charge interactions. In Fig.\,\ref{gaptc}(c) and (d) we show $V_{\mathbf{q},l=0}^{(\mathrm{s})}$ and $V_{\mathbf{q},l=0}^{(\mathrm{c})}$, respectively. In both panels different values for $U$ are represented by the choice of colors, while solid and dashed lines correspond to $g_0=0\,\mathrm{meV}$ and $g_0=180\,\mathrm{meV}$, i.e., the absence or presence of EPI. We observe that the kernel from charge fluctuations is approximately one order of magnitude smaller than the spin-fluctuations part. Furthermore, as we increase the 
$U$ parameter, both interactions grow throughout the entire BZ, while $V_{\mathbf{q},l=0}^{(\mathrm{s})}$ tends to diverge around the nesting wave vector $\mathbf{q}\sim(\pi,\pi)$. 

It is important to note that the spin-fluctuations kernel enters repulsively in the equation for the superconducting order parameter, compare Eq.\,(\ref{phi}), which means that the interaction at exchange wave vector $\mathbf{q}\sim(\pi,\pi)$ promotes a $d$-wave symmetry of the gap. However, in the absence of EPI the repulsive kernel around $\mathbf{q}  \sim(0,0) = \Gamma$ does not support this symmetry, which leads to phase oscillations {during the self-consistent iteration cycles,}
and, eventually, to $\Delta=0$. This behavior is analogous to that found in recent theoretical studies on Fe-based superconductors \cite{Yamase2020,Schrodi2020_3,Schrodi2020_4,Schrodi2021_3}. In the presence of EPI we find a stable superconducting solution to the Eliashberg equations, which reduces the spin and charge kernels slightly throughout the entire BZ (see Figs.\,\ref{gaptc}(c) and (d)). Most importantly, we see a significant decrease in $V_{\mathbf{q},l=0}^{(\mathrm{s})}$ around $\mathbf{q}=(0,0)$, hence the system adjusts the repulsive small-$\mathbf{q}$ contributions, thereby promoting the Cooper pair formation.

\section{Discussion and Conclusions}

We have introduced an efficient method to study self-consistently multichannel superconductivity, by treating vertex-corrected EPI and SFs/CFs {together in one computational formalism.}
Our methodology opens a pathway for studying other anomalous
superconductors in a systematic way, which can be achieved via materials' specific choices of the phonon frequency, electron energy dispersion, and coupling strengths for both mediators of superconductivity. The here-introduced non-interacting state approximation for the electron-phonon vertex function is of utter importance for making the calculations numerically 
feasible.

At this point it is appropriate to summarize the limitations of the here-developed computational approach. The non-interacting state approximation for the electron-phonon vertex is justified for the studied cuprate system (see Appendix \ref{Appendix}), but this might not be the case for other non-adiabatic superconductors. Vertex-corrected Eliashberg theory that includes second-order scattering processes is expected to be valid for weakly non-adiabatic systems, but higher order scattering processes might well play a role for strongly non-adiabatic systems. Another step towards a more complete picture  would be  the  inclusion  of  phonon  renormalization  effects and  the  Coulomb  repulsion,  which  were  both  assumed here to be included in our parameters. A renormalization of the magnon spectrum due to electron-phonon coupling is not accounted for. Also, the spin and charge fluctuations are treated in a single orbital model, but for a more realistic description a multiorbital model might be required  \cite{Takimoto2004}. On a more general note, it deserves to be mentioned that the question about effects due to vertex corrections to the purely electronic interactions remains unanswered, which is of particular importance due to the absence of an analogue of Migdal's theorem for SFs \cite{Hertz1976}.

Applying our methodology to a 
cuprate model system, we have found that the EPI is responsible for the gap magnitude and high critical temperature, while the SF part of the electronic interaction suppresses superconductivity.
Both interactions support the $d$-wave symmetry order parameter via repulsive 
coupling {at large wave-vectors}. We found that repulsive small momentum wave-vector contributions from the SF kernel are responsible for the reported suppression of $T_c$. Since a similar behavior has been encountered in Fe-based compounds \cite{Yamase2020,Schrodi2021_3}, it might be a generic effect in high-temperature superconductors.

Our cuprate model system has roughly the correct characteristics of the family of compounds that we want to investigate, i.e., the prototypical FS of electron-doped cuprates. Additionally, the phonon frequency and EPI coupling strength $\lambda_m$ are in realistic ranges \cite{Lanzara2001,Iwasawa2008}. For the SF part of the interaction, a model similar to ours has been shown to be accurate for many characteristics of the interacting state \cite{Moriya2003}.
Our results suggest therefore that electron-phonon interaction can well be the prime mediator of unconventional superconductivity in high-$T_c$ cuprates.

\begin{acknowledgments}
This work has been supported by the Swedish Research Council (VR), the R{\"o}ntgen-{\AA}ngstr{\"o}m Cluster,  and the Knut and Alice Wallenberg Foundation (grant No.\ 2015.0060).
The calculations were enabled by resources provided by the Swedish National Infrastructure for Computing (SNIC) at NSC Linköping, partially funded by the Swedish Research Council through Grant No.\ 2018-05973.
\end{acknowledgments}

\appendix

\section{Accuracy of vertex approximation} 
\label{Appendix}

\begin{figure}[h!]
	\centering
	\includegraphics[width=1\columnwidth]{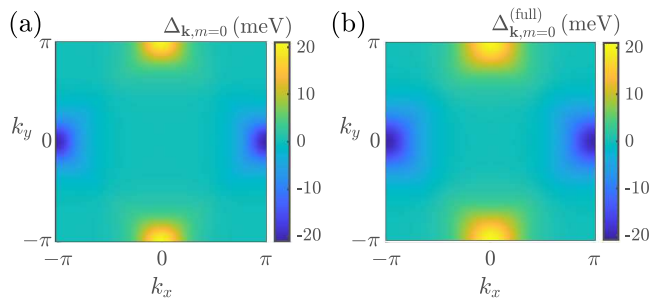}
	\caption{Self-consistently calculated superconducting gap function at zero frequency, using $\Omega=50\,\mathrm{meV}$, $g_0=175\,\mathrm{meV}$, and $T=80\,\mathrm{K}$. (a) The gap function computed using the non-interacting state approximation for the vertex. (b) The complete solution computed without any approximation.}
	\label{benchmark}
\end{figure}

Here we briefly give a justification of the non-interacting state approximation for the vertex, which is employed in 
Sec.\,\ref{Results}.
Since this only concerns the EPI, we set $U=0$ and therefore neglect any influence of SFs and CFs. We use the phonon frequency $\Omega=50\,\mathrm{meV}$ and set $g_0=175\,\mathrm{meV}$. Further, we choose the relatively high temperature $T=80\,\mathrm{K}$ so as to reduce the required number of Matsubara frequencies. In Fig.\,\ref{benchmark}(a) we show our result for $\Delta_{\mathbf{k},m=0}$, which is obtained by using the formalism described in the main text, i.e., under the non-interacting state approximation for the vertex function, compare Eq.\,(\ref{nonint}). As already described in the main text, we find a $d$-wave symmetric gap function, here with a maximum value of $\max_{\mathbf{k}}\Delta_{\mathbf{k},m=0}=21.1\,\mathrm{meV}$. The calculated maximum of the mass renormalization is $\max_{\mathbf{k}}Z_{\mathbf{k},m=0}=1.72$.

We can compare these findings to results obtained from the full formalism without approximations, i.e., using Eq.\,(\ref{vertex}) as was done in Refs.\,\cite{Schrodi2020_2,Schrodi2021}. To distinguish such results from our current theory, we use the label `(full)'. For these results, shown in Fig.\,\ref{benchmark}(b), we observe the same $d$-wave symmetry of the superconductivity order parameter, with remarkably similar extremum $\max_{\mathbf{k}}\Delta_{\mathbf{k},m=0}^{(\mathrm{full})}=20.5\,\mathrm{meV}$. The same holds for the electron mass renormalization, which now takes on a maximum value of $\max_{\mathbf{k}}Z^{(\mathrm{full})}_{\mathbf{k},m=0}=1.70$. From these outcomes it is evident that the non-interacting state approximation to the vertex is reliable for the cuprate model system used in the current work.

\bibliographystyle{apsrev4-1}

%

\end{document}